\documentclass[twocolumn,final,fleqn]{svjour2}
\usepackage{pslatex}
\usepackage{latexsym}
\usepackage{amsmath,amssymb}
\usepackage{graphicx}
\usepackage{dcolumn}
\usepackage{bm}
\usepackage[numbers]{natbib}
\journalname{Granular Matter}

\begin{document}
\title{Monitoring Three-Dimensional Packings in Microgravity}
\author{Peidong Yu \and Stefan Frank-Richter \and Alexander B\"orngen 
\and Matthias Sperl}
\institute{Institut f\"ur Materialphysik im Weltraum,
Deutsches Zentrum f\"ur Luft- und Raumfahrt, 51170 K\"oln, Germany
\\\email{matthias.sperl@dlr.de}
}
\date{\today}

\maketitle 
\begin{abstract}

We present results from experiments with granular packings in three 
dimensions in microgravity as realized on parabolic flights. Two different 
techniques are employed to monitor the inside of the packings during 
compaction: (1) X-ray radiography is used to measure in transmission the 
integrated fluctuations of particle positions. (2) Stress-birefringence in 
three dimensions is applied to visualize the stresses inside the packing. 
The particle motions below the transition into an arrested packing are 
found to produce a well agitated state. At the transition, the particles 
lose their energy quite rapidly and form a stress network. With both 
methods, non-arrested particles (rattlers) can be identified. In 
particular, it is found that rattlers inside the arrested packing can be 
excited to appreciable dynamics by the rest-accelerations (g-jitter) 
during a parabolic flight without destroying the packings. At low rates of 
compaction, a regime of slow granular cooling is identified. The slow 
cooling extends over several seconds, is described well by a linear law, 
and terminates in a rapid final collapse of dynamics before complete 
arrest of the packing.

\end{abstract}

%%%
\section{Introduction}\label{sec:intro}

Experiments with granular matter in microgravity allow access to regions 
in control-parameter space that are otherwise not accessible. Microgravity 
prevents the sedimentation of a loose non-agitated granular assembly and 
hence enables the long-term study of such states. For agitated granular 
matter, experiments in microgravity can reduce the inhomogeneity of driven 
states; and for particles in contact, the absence of gravity eliminates 
the pressure gradient in the packings. To what extent these goals can be 
realized in a specific experiment depends largely on the quality of the 
microgravity conditions found on specific platforms. Experiments have been 
performed for granular gases \cite{Falcon1999,Harth2013,Sack2013} as well 
as dense systems under shear \cite{Murdoch2013,Murdoch2013a}. In the 
following sections, it will be shown how the microgravity environment of a 
parabolic flight can be utilized for investigating granular packings. 
Results will be elaborated for X-ray radiography as well as 
stress-birefringence in three dimensions.

\begin{figure}[hbt]\begin{center}
\includegraphics[width=.95\columnwidth]{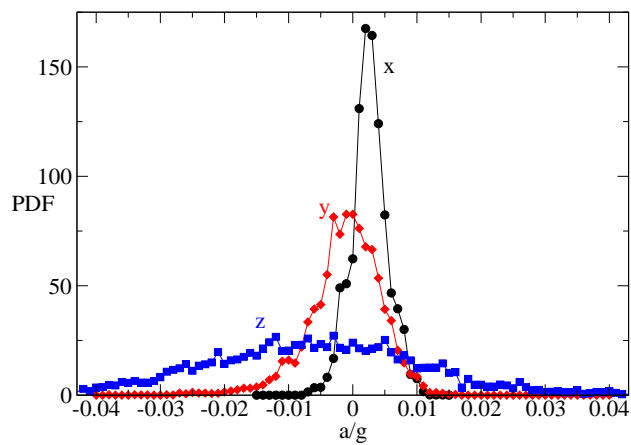}
\caption{\label{fig:jitter}
Distribution of rest accelerations averaged over a typical single 
parabola of 22 seconds on the third day of parabolic flight campaign 
DLR-22. Data are shown for the $x$- (circles, forward direction of the 
airplane), $y$- (diamonds, wing-to-wing direction of the airplane), and 
$z$-directions (squares, floor-to-ceiling direction of the airplane).
}\end{center}
\end{figure}

%%%
\section{Microgravity}\label{sec:micro}

Microgravity environments are typically hard to obtain and require years 
of preparation. In contrast to experiments in space, parabolic flight 
campaigns offer a reasonably frequent opportunity to perform experiments 
under microgravity conditions. While not offering the best microgravity 
quality in terms of rest-accelerations, cf. discussion below, parabolic 
flights can help to test phenomena that depend on a distinction between 
top and bottom. One such phenomenon is convection. For a granular system 
under shear, convection was found perpendicular to the direction of shear 
along the direction of gravity \cite{Khosropour1997}. On a parabolic 
flight however, it was observed recently, that in the absence of a 
distinction between top and bottom such convection disappears 
\cite{Murdoch2013}.

The limitation of such experiments on parabolic flights is the presence of 
rest-accelerations -- called $g$-jitter -- which drastically restrict the 
time the particles can stay in a granular gas without being collectively 
driven against the container walls within around a second. For dense 
granular matter, the $g$-jitter imposes a minimum necessary confinement 
for keeping granular packings confined.

Figure~\ref{fig:jitter} shows the distribution of rest accelerations on a 
parabolic flight, averaged over a single flight day. The $x$-direction is 
defined from the tail to the front of the plane, the $y$-direction is from 
the left to the right wing when looking from the cabin to the cockpit, and 
the $z$-direction points from the floor to the ceiling of the cabin. Given 
the rather uncontrolled nature of the rest-accelerations, it is remarkable 
how well they follow reasonable distributions. The full width of the 
distributions at half maximum in units of $g$ is 0.005 for the 
$x$-direction, 0.01 for the $y$-direction, and 0.04 for the $z$-direction. 
In addition to the width of the distributions showing rather large 
qualitative differences, also the maximum values in $x$- and 
$z$-directions show deviations from zero, $a_x^0/g \approx 0.0025$ 
(forward bias) $a_z^0/g \approx -0.012$ (downward bias). The $y$-direction 
is on average symmetric. Data for a single parabola typically look similar 
to Fig.~\ref{fig:jitter} while being somewhat variable between individual 
parabolas.

Rather than trying to avoid the influence of the rest-accele\-rations, in 
the following experiments the $g$-jitter is utilized for providing 
agitation for dense granular systems.

%%%
\section{X-Ray Radiography}\label{sec:xray}

\begin{figure}[hbt]\begin{center}
\includegraphics[width=.95\columnwidth]{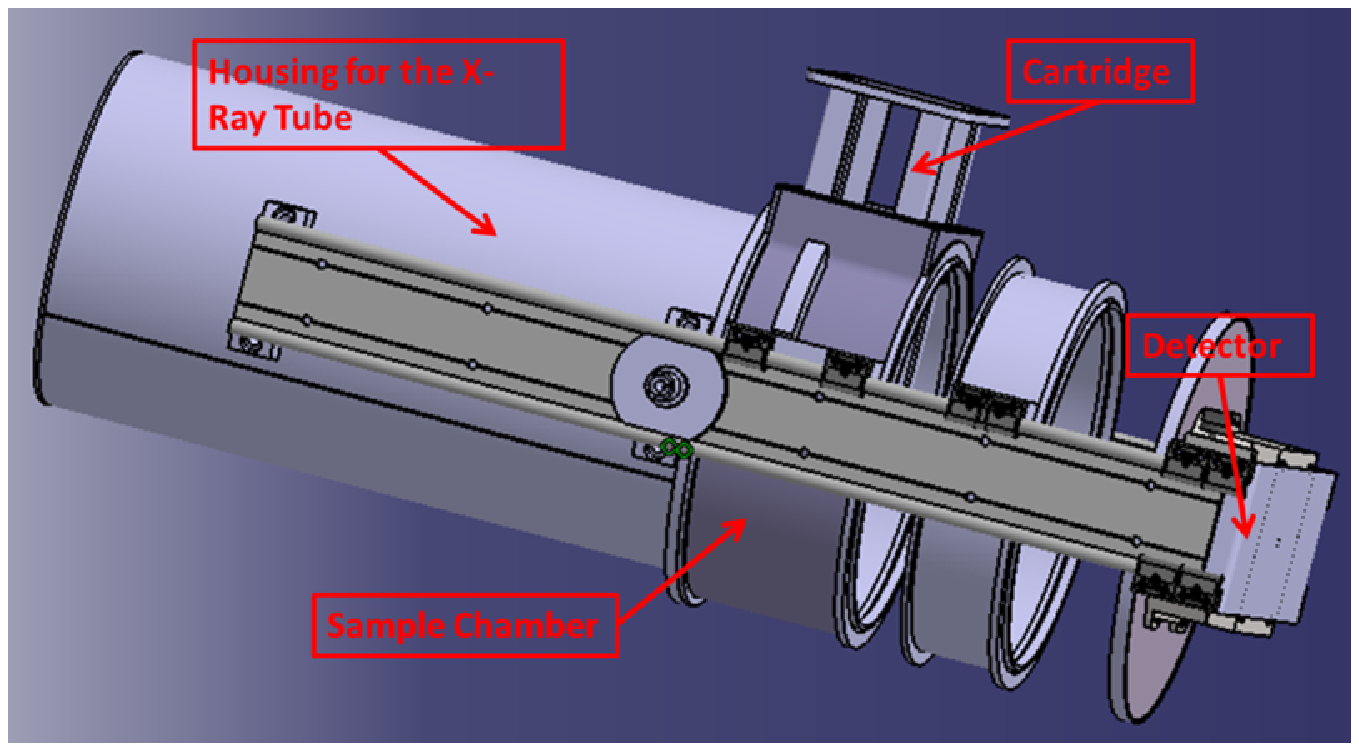}
\includegraphics[width=.95\columnwidth]{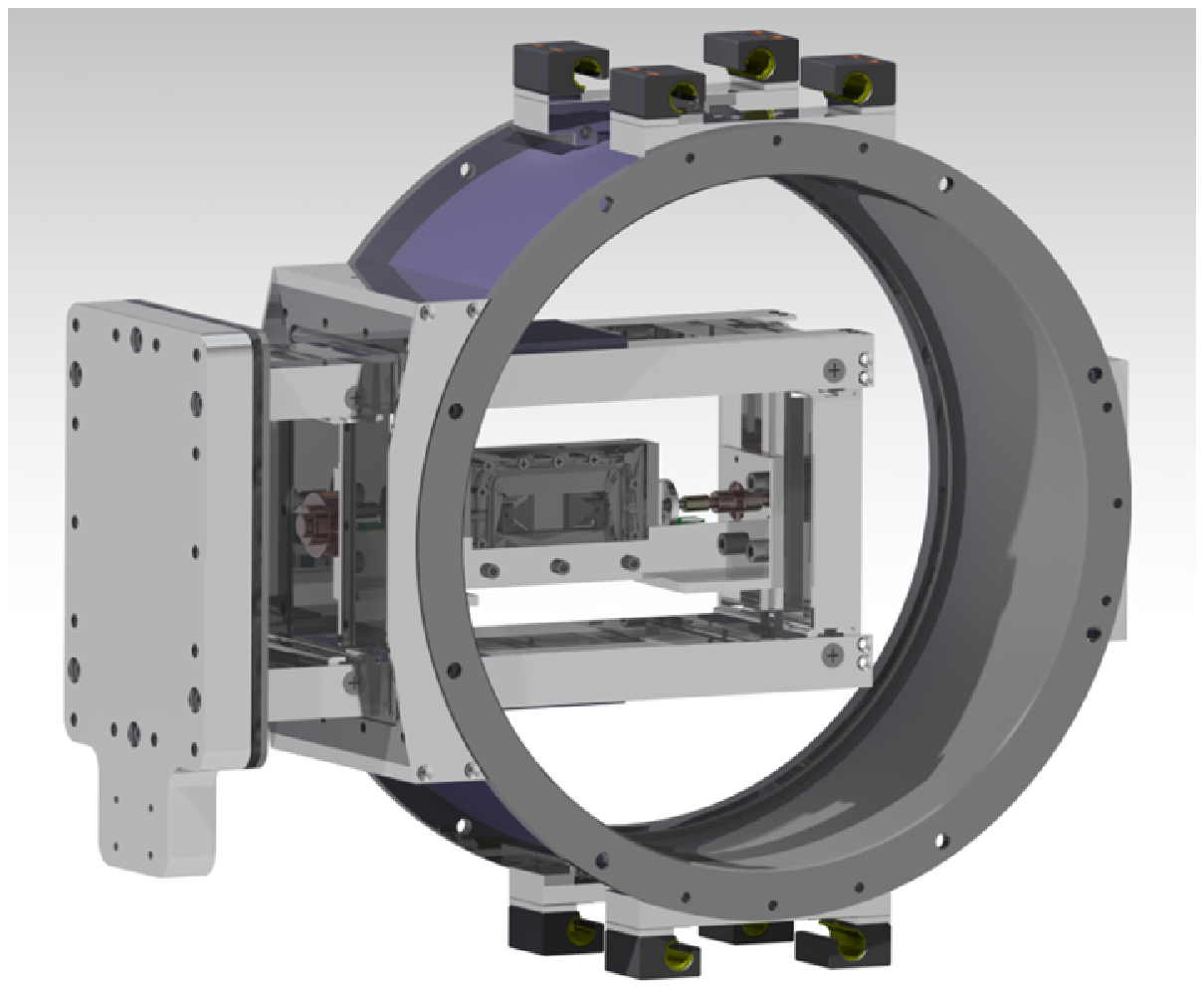}
\includegraphics[width=.95\columnwidth]{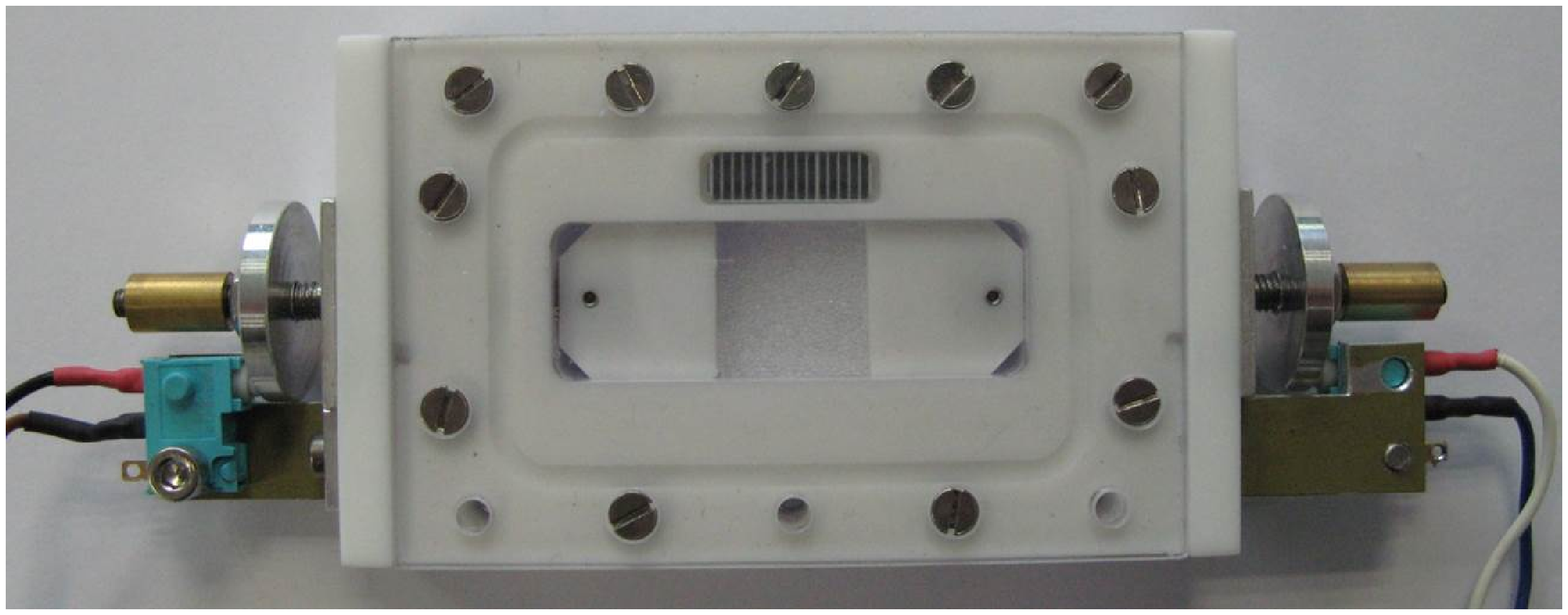}
\caption{\label{fig:setupXray}Experimental setup for the parabolic flight 
(DLR-22, April 2013) for X-ray radiography. The top panel shows a 
schematic view from left to right of the X-ray tube, the experiment 
chamber with sample cartridge, a spacer ring, and the detector. The 
central panel exhibits the sample chamber with the replaceable cartridge 
for granular experiments which is shown in the photograph of the bottom 
panel. The granular cartridge has two motorized pistons of cross-section 
15mm$\times$5mm and an X-ray ruler with a mm-scale on top.
}\end{center}
\end{figure}

The use of X-ray illumination facilitates the visualization of otherwise 
optically opaque samples. The simplest use of a combination of an X-ray 
source and a detector is by recording the transmission images after 
absorption from the sample in a radiography setup. X-ray radiography has 
been used to investigate hopper flow of sand \cite{Baxter1989} as well as 
the dynamics of granular matter in fluidized beds 
\cite{Grohse1955,Rowe1972,Yates2002}. The addition of tomography, i.e. 
rotating the still sample for multiple transmission images, allows the 
reconstruction of packings \cite{Aste2006,Jerkins2008}.

The aim of the present study using X-ray radiography is to investigate the 
compaction of a granular assembly into a dense packing. On ground the 
compaction is dominated by gravity-induced sedimentation and takes place 
rather rapidly within a fraction of a second and also comparably violently 
with shock waves traveling through the system \cite{Son2008}. In 
microgravity, the energy loss is still driven by interparticle collision 
but the rapid sedimentation is replaced by the compaction from the 
container walls which is chosen here to be rather moderate in speed.

Figure~\ref{fig:setupXray} shows the setup of the radiography device. The 
source produces a divergent X-ray beam that irradiates a sample before 
being registered by the detector (CCD-/COOL-1100XR) with pixel size 
9$\mu$m$\times$9$\mu$m recording with a resolution of 2008$\times$1340 
pixels and 16-bit depth at 4~fps (frames per second). The placement of the 
sample between source and detector as well as their overall distance 
determines the magnification. Additional spacer rings can be used to 
increase the possible magnification. In the following, a magnification 
factor of 2 was chosen. The actual sample cell is placed in a sample 
chamber in the form of a replaceable cartridge. In addition to changing 
granular samples easily, also samples other than granular matter can be 
used with the device. As seen in the bottom panel of 
Fig.~\ref{fig:setupXray}, the granular sample cell contains the sample 
material inside a rectangular volume that can be changed by pistons on two 
sides. An X-ray ruler with a millimeter scale is used to calibrate the 
volume and hence the packing fraction of the experiments.

\begin{figure}[hbt]\begin{center}
\includegraphics[width=.495\columnwidth]{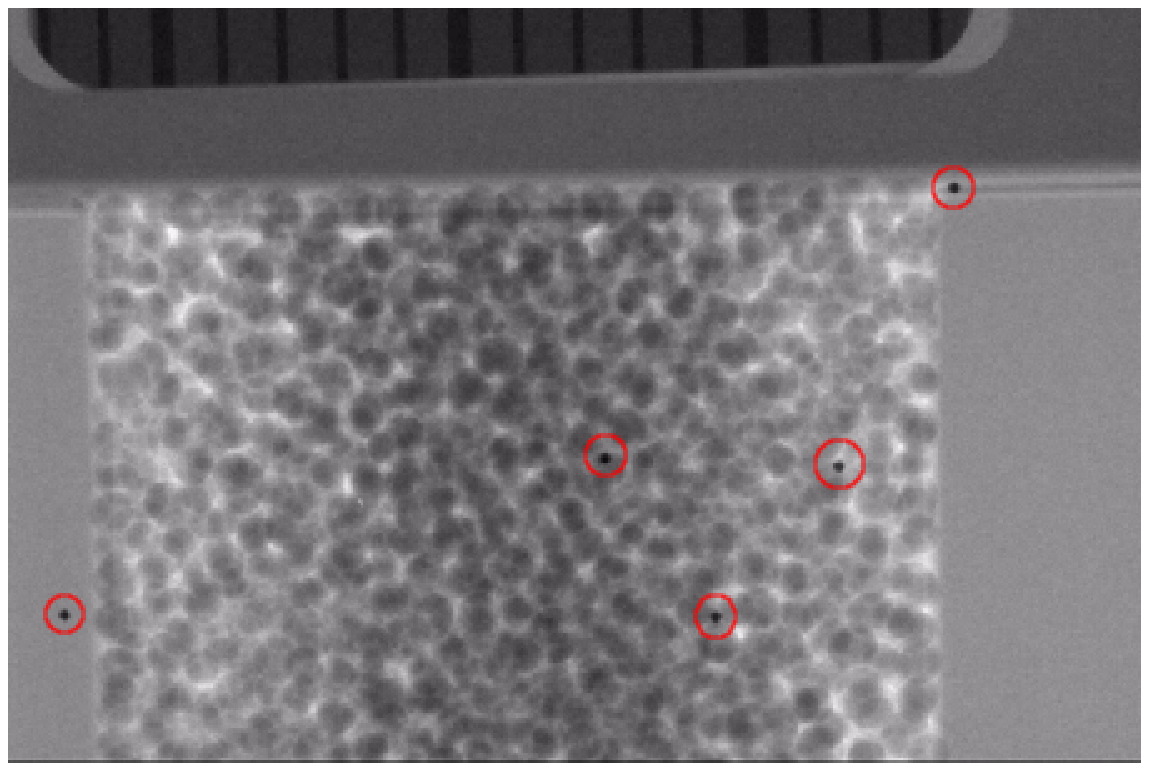}
\includegraphics[width=.495\columnwidth]{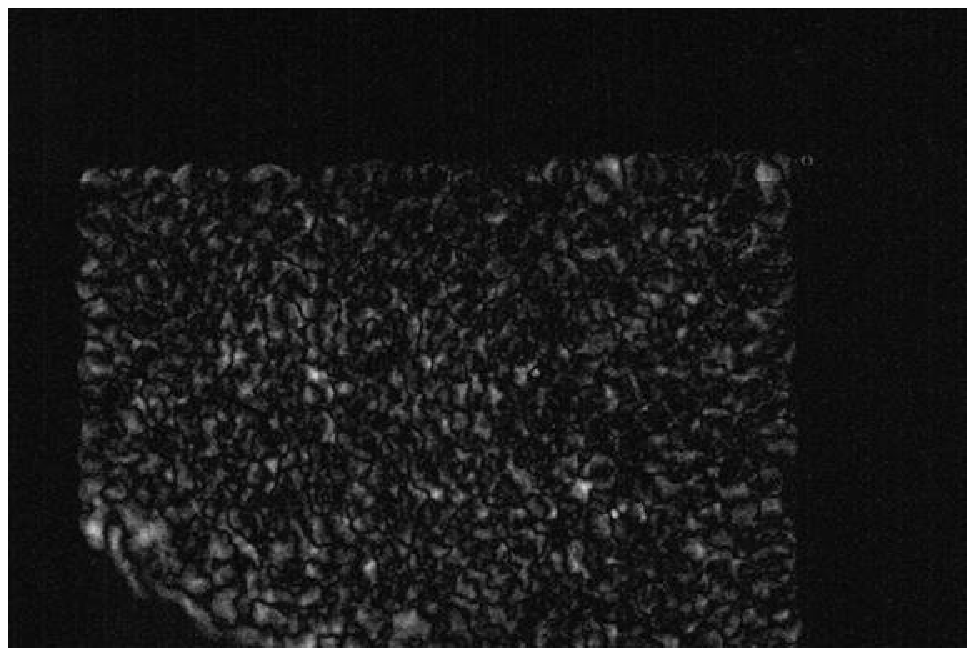}
\\
\includegraphics[width=.495\columnwidth]{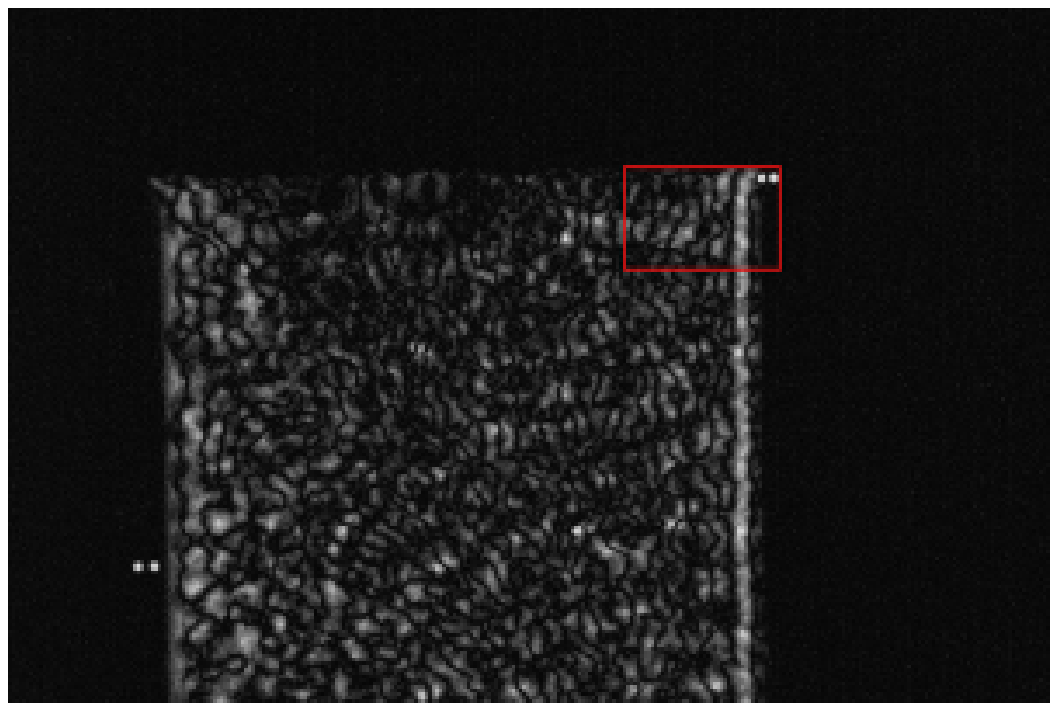}
\includegraphics[width=.495\columnwidth]{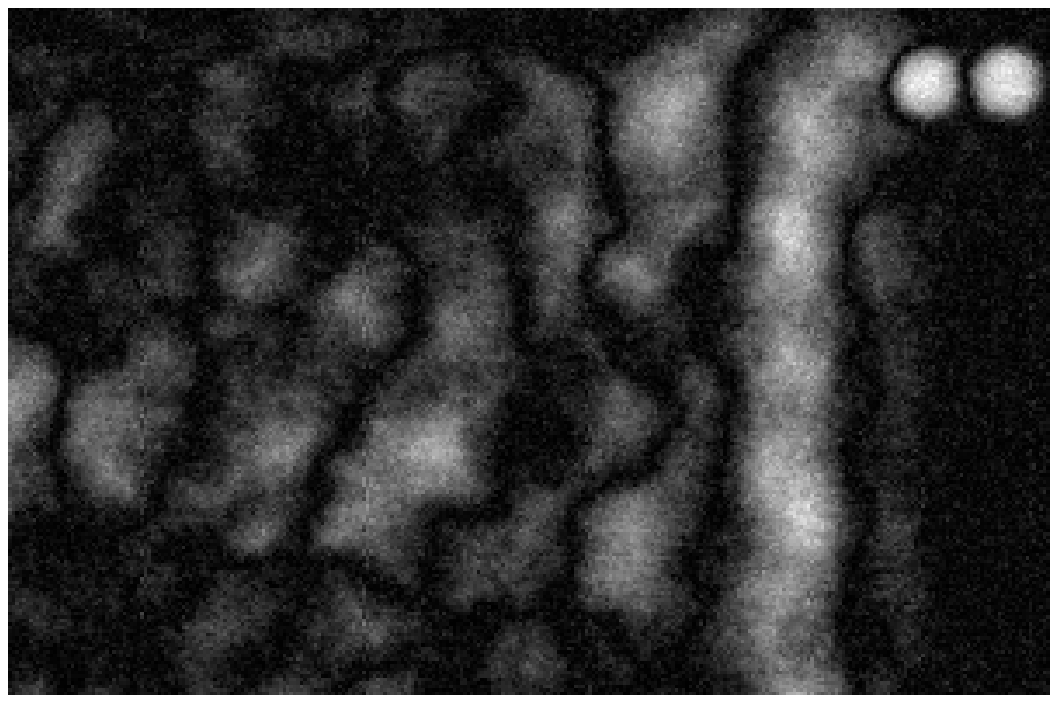}
\\
\includegraphics[width=.495\columnwidth]{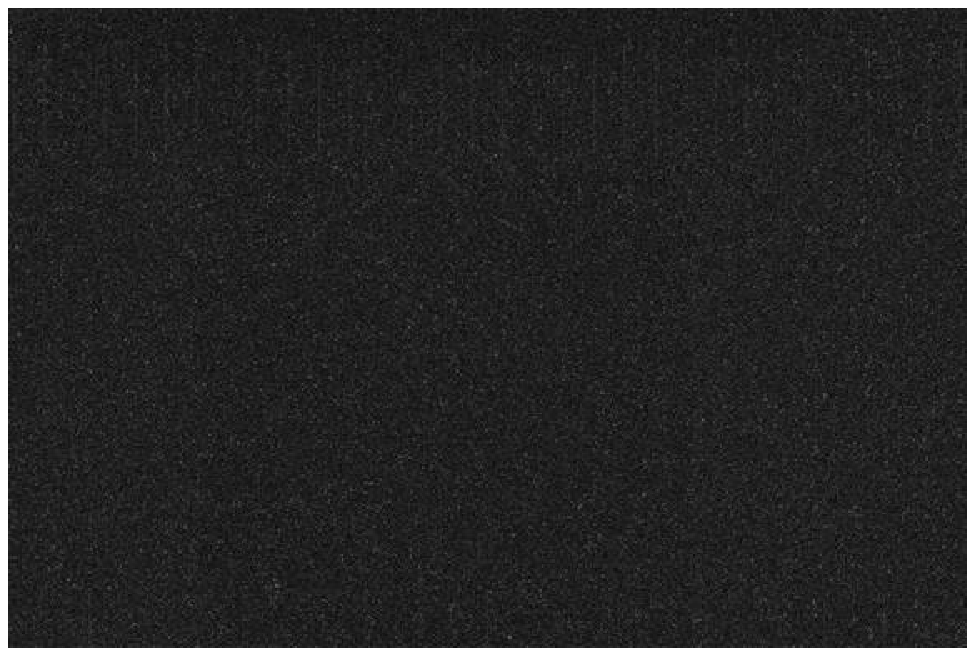}
\includegraphics[width=.495\columnwidth]{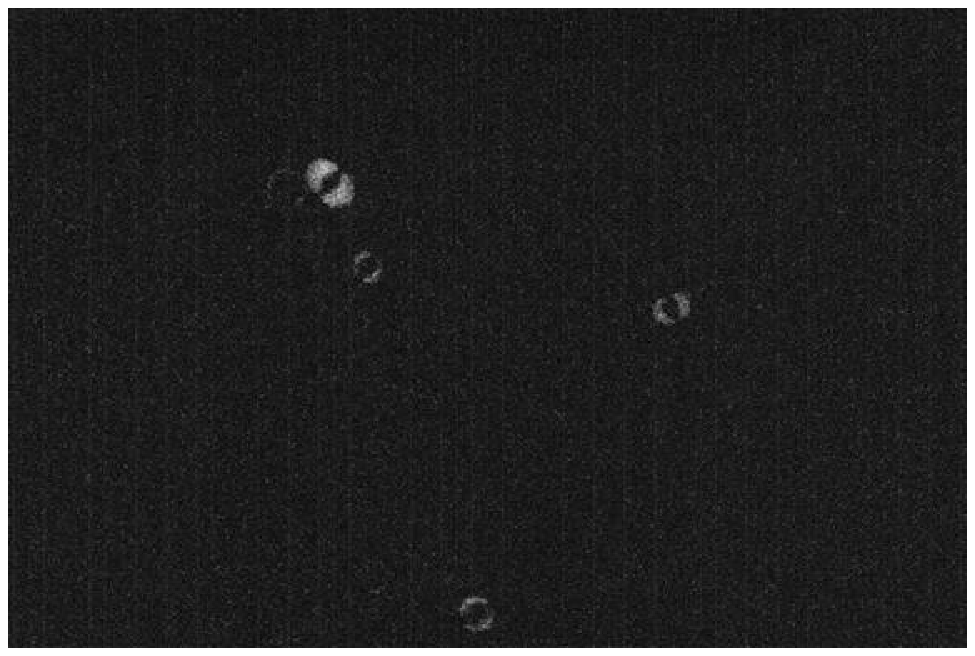}
\caption{\label{fig:P1}Radiography images from parabola number 1. The 
original transmission image (upper left) shows the pile of glass particles 
(diameter 500$\mu$m) in the 2-g phase before the microgravity experiment. 
Gravity acts perpendicular to the plane of the image. Darker particles 
singled out by red circles are steel particles (diameter 200$\mu$m) acting 
as tracers. The difference image (upper right) shows the motion between 
two successive frames due to g-jitter at the beginning of compaction 
(recording at four frames per second). A similar image (middle left) shows 
the differences immediately after motion of the pistons together with a 
rectangular frame for the enlarged selection shown on the next difference 
image (middle right). The fourth difference image (lower left) shows the 
absence of detectable motion after compaction and cooling of the arrested 
sample. The final difference image (lower right) illustrates the motion of 
four rattler particles at the transition from the 0g to the 2g-phase.
}\end{center}
\end{figure}

The device described above was used in the parabolic flight campaign 
DLR-22 in April 2013. The orientation of the X-ray beam was chosen in the 
$z$-direction of the airplane, so the largest dimensions of the sample 
cell were in the $x$-$y$-plane of the aircraft where the least overall 
bias of the $g$-jitter could be expected. The sample volume was filled 
with around 8000 glass particles of diameter 500$\mu$m (estimated 
coefficient of restitution $\varepsilon \approx 0.7$). Tracer particles of 
diameter 200$\mu$m were added to have access to individual particle 
trajectories. These particles were made from steel to ensure good contrast 
which is seen in the first panel of Fig.~\ref{fig:P1}. The choice of 
tracer particles much smaller than the particles of the host system was 
motivated by the resolution limitations in both space and time: Smaller 
particles are more likely to be rattlers, i.e. show appreciable motion 
even inside an arrested state. The volume was filled with particles on 
ground and compacted with the pistons to form a stable packing without 
deforming the particles. Afterwards the pistons were retracted 
symmetrically and left the granular particles in a pile as seen in the 
upper left panel of Fig.~\ref{fig:P1} with more particles in the center 
than closer to the pistons. This asymmetry vanishes immediately after 
entering the microgravity phase where the $g$-jitter redistributes the 
particles homogeneously in the sample volume.

After agitation of the granular particles by $g$-jitter, the system was 
slowly compressed by the pistons from a packing fraction of around 
$\varphi = 0.43$ until the arrested state around $\varphi=0.6$ was 
reached. The reported packing fractions are calculated from dividing the 
volume of the particles by the full available volume of the test cell. For 
the packed state we estimate the deviation of the true bulk packing 
fraction from the nominal one as follows: We subtract from the particle 
volume the sum of the half spheres of a completely covered layer of 
particles at the walls. From the cell volume we subtract the corresponding 
sum of half-cubes. The resulting boundary-corrected value for the packing 
fraction at the arrested state, $\varphi = 0.6$, is found at 
$\tilde\varphi = 0.615$, i.e. a deviation of 2.5\% for the bulk value 
inside the sample. Since this correction is not reasonable for more dilute 
assemblies down to nominal packing fraction of 0.43, the nominal values 
are reported in the following. Even accounting for the outlined boundary 
correction, the arrested sample does not reach values for the 
packing fraction commonly reported for random-close packing of around 
$\varphi = 0.64$. The lower packing fraction at the arrested state in our 
samples is explained by the comparably high friction among the particles. 

The difference image in the upper right panel of Fig.~\ref{fig:P1} shows 
the absolute intensity variation from one frame to the successive one and 
hence characterizes the overall motion across the sample. It is found that 
the particles at the initial volume are quite well agitated. The volume of 
particles in that difference image is distinguished well from the 
container walls which do not move and appear black plus some noise. The 
middle panels of Fig.~\ref{fig:P1} show the motion immediately after 
compression by the pistons which is visible by the two trapped tracer 
particles on the lower-left and upper-right corners. While on the right 
wall a whole layer of particles is displaced together, on the left wall 
the energy input yields a more random pattern. This difference is not very 
surprising as the particle density at both walls is not necessarily the 
same before the particles are packed densely. A rectangular frame in the 
middle left panel indicates an area in the full test cell that is shown 
magnified by a factor of seven in the middle right panel. It is clear from 
the enlarged image that in the setup individual tracer particles can be 
resolved.

Once the final close-packed volume is reached, the motion in the sample 
cell vanishes as seen by the completely dark difference image in the lower 
left. Container and particle packing are then indistinguishable. The final 
difference image in Fig.~\ref{fig:P1} shows the observations at the 
transition from microgravity to 1.8$g$ at the end of a parabola: As both 
new and old position of a particle show up brightly, four individual 
particles can be identified as moving on the timescale of a quarter 
second. We interpret these as rattlers that have lost all their energy 
during cooling inside the packing and are now pulled downwards by the 2$g$ 
acceleration.

\begin{figure}[hbt]
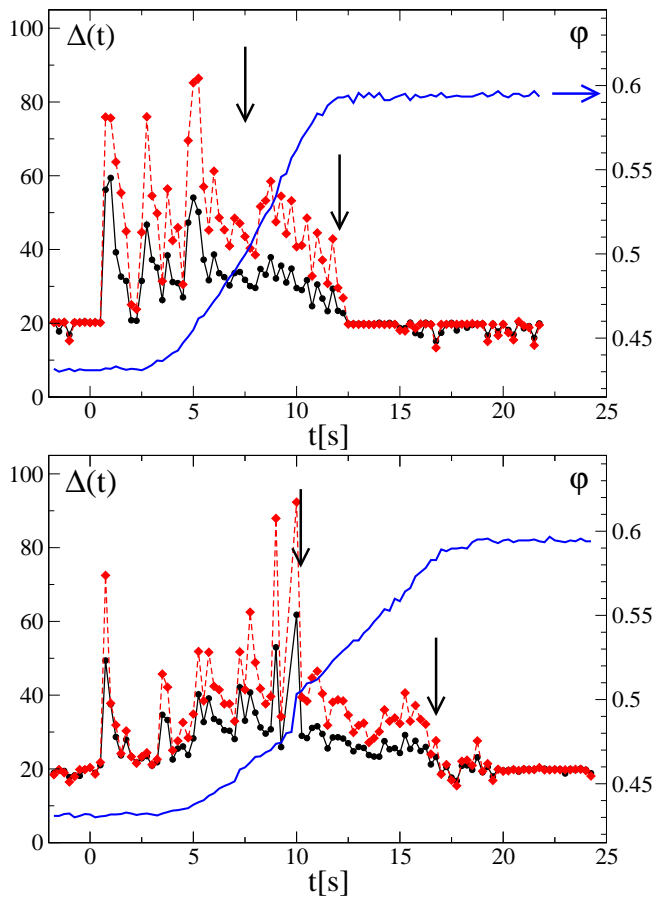

\includegraphics[width=.99\columnwidth]{Diffmean_P10.eps}
\includegraphics[width=.99\columnwidth]{Diffmean_P1.eps}
\caption{\label{fig:cooling}Dynamics of the granular particles during the 
slow compaction process. The plots display the overall brightness of 
successive difference images $\Delta(t)$ on the left axes over time during 
the microgravity phase for a representative compaction run within 10 
seconds (parabola 10, upper panel) and a run within 13 seconds (parabola 
1, lower panel). The respective right axes display the evolution of the 
packing fraction for the full curve. The two curves for $\Delta(t)$ 
show the average over the full sample (filled circles) and the center of 
the cell without boundaries (diamonds). Vertical arrows indicate a region 
of slow cooling (see text).}
\end{figure}

The time evolution of the brightness in the difference images can serve as 
an estimate of the granular system's kinetic energy and hence the decrease 
in brightness signals granular cooling. This evolution of the brightness is 
shown in Fig.~\ref{fig:cooling}. The brightness of the difference images 
$\Delta(t)$ is defined by the averaged greyvalue per pixel over a region of 
interest. The region of interest is taken either for the entire probe-cell 
volume (with the trade-off of including the pistons for the later part of 
the compaction) shown as diamonds as well as over only the central quadratic 
region filled with particles after compaction without any boundaries shown 
by the filled circles (with the trade-off of missing some particles close to 
the walls at the earlier part). Both definitions of the region of interest 
yield no qualitative difference in the observed data, so it seems both 
definitions capture the particle dynamics reasonably well and the dynamical 
features are dominated by the behavior in the bulk. The origin of the time 
scale is set to the beginning of the 0$g$ phase. The compaction is seen by 
the evolution of the packing fraction over time. The overall packing 
fraction is reduced by $\Delta\varphi/\Delta t = 0.017/s$ for compaction in 
10s and by $\Delta\varphi/\Delta t = 0.013/s$for compaction in 13s, 
respectively. For those slow compaction rates, data from 10 parabolas was 
used. Similar five runs have been obtained for a fast compaction rates of 
$\Delta\varphi/\Delta t = 0.04/s$ which is not shown in the figure but 
discussed below.

For slow compaction, at both reported compaction rates the reproducible 
observations can be summarized as follows. 

(1) Throughout all the runs, both for the beginning when particles are at 
rest in 1.8$g$ and at the end of compaction when still in a noisy 0$g$ 
environment, the background value is always $\Delta_0 = 20$. There is no 
observable drift in $\Delta_0$ and in the $\Delta(t)$ over different runs. 
Faster overall motion of the particles as apparent from the original 
images is reflected in a higher amplitude of $\Delta$.

(2) At the start of 0$g$, the system is shaken strongly and exhibits 
strong fluctuations in $\Delta(t)$ seen by the large peaks in both panels 
of Fig.~\ref{fig:cooling} on the respective left sides. The fluctuations 
are not affected by the compaction which is setting in after a few seconds 
in 0$g$.

(3) Around $\varphi = 0.5$ (indicated by vertical arrows in 
Fig.~\ref{fig:cooling}) fluctuations are dampened and the evolution of 
$\Delta(t)$ suggests a regime a granular cooling. This cooling regime was 
found for 10 out of 11 runs with slow compaction. For the single exception 
the pistons got stuck and snapped before a cooling regime can be 
identified in the data. A reminiscence of that stick-slip piston behavior 
can be seen around 10s in the lower panel of Fig.~\ref{fig:cooling} in the 
curve for the packing fraction.

(4) The cooling regime shows up similarly for both definitions of a region 
of interest; the more restricted region of interest (diamonds) is used for 
the quantitative analysis in the following. The slow cooling can be 
described by a linear law $\Delta(t) -\Delta_0 = \tilde\Delta \gamma t$ 
where $\tilde\Delta$ describes the overall amplitude, i.e. the equivalent of 
granular temperature, at the beginning of the cooling. For the amplitude we 
obtain $\tilde\Delta = 40$ for the upper panel in Fig.~\ref{fig:cooling} and 
$\tilde\Delta = 30$ for the lower panel. Parameter $\gamma$ describes a 
normalized cooling rate that turns out to be well reproducible across all 10 
parabolas for slow cooling with no significant difference for different 
compaction rates: $\gamma = 0.13\pm0.02/s$.

(5) The linear regime for slow cooling is terminated upon reaching the 
final packing fraction by a fast cooling regime where within around $1s$ 
the complete dynamics comes to rest, i.e. $\Delta(t) = \Delta_0$. The 
limited time resolution of the data does not allow a more quantitative 
statement, but the fast cooling regime is always identified clearly, the 
linear regime for slow cooling does not extend all the way to $\Delta_0$. 
After the fast cooling regime, the sample is arrested. Note that the 
appearance of rattlers as seen in Fig.~\ref{fig:P1} is not visible above 
the noise level in $\Delta(t)$.

\begin{figure}[hbt]
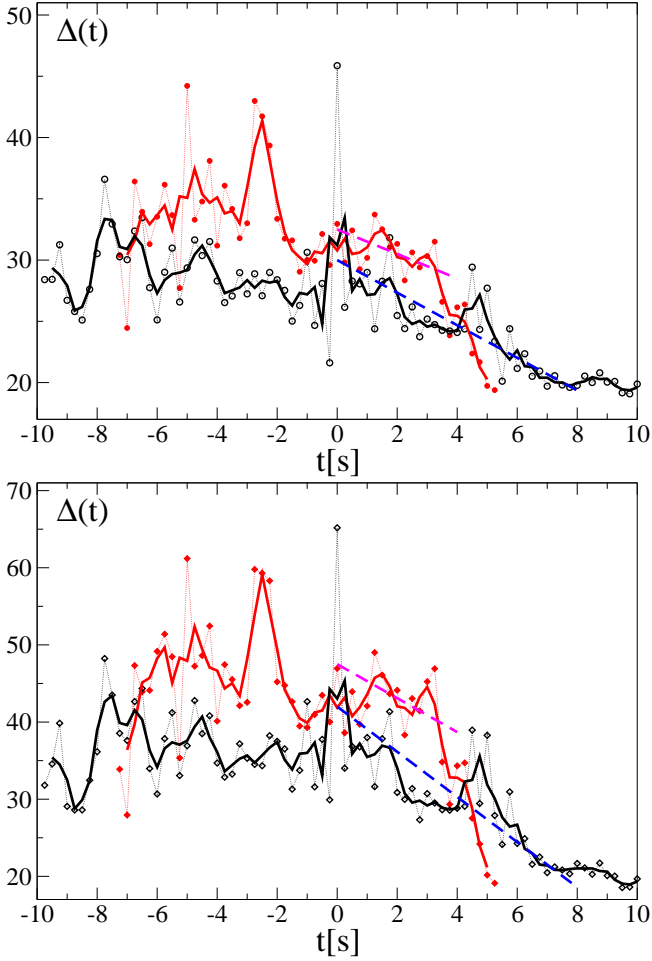

\includegraphics[width=.99\columnwidth]{avgallpix.eps}
\includegraphics[width=.99\columnwidth]{avgselpix.eps}
\caption{\label{fig:avg}Averages over the particle dynamics during the slow 
compaction process evaluated for all pixels (top panel) and the selected 
region of pixels (bottom panel). The origin of time is set to the time when 
$\varphi = 0.5$ for each run. The open symbols in both panels represent the 
average over parabolas P0 to P5 (compaction rate $\Delta\varphi/\Delta t = 
0.013/s$) while the full symbols show results from parabolas P6 to P10 
(compaction rate $\Delta\varphi/\Delta t = 0.017/s$). Full curves are 
corresponding running averages in time over 0.5s, i.e. the average of three 
data points. Dashed straight lines display the linear laws $\Delta(t) = 
\tilde\Delta\gamma t$.
}
\end{figure}

Observations (1) to (5) as elaborated above are found for all realizations 
of slow compaction for 10 parabolas. In particular, the limit of $\varphi = 
0.5$ where fluctuations become smaller and cooling sets in, is reproducible 
across the available data. If the compaction is around four times faster as 
investigated for additional five parabolas, no such limit exists and no such 
regime of slow cooling can be identified. Also, in Fig.~\ref{fig:cooling} 
one observes that the range of validity for the linear law shrinks from 6.5s 
for compaction rate $\Delta\varphi/\Delta t = 0.013/s$ to 4.5s for 
$\Delta\varphi/\Delta t = 0.017/s$. Hence, we conclude that the existence of 
a slow cooling regime depends on the balance between energy input (from 
$g$-jitter and the compaction process) and the rate of dissipation (given by 
$\varepsilon$) and can be tuned by the rate of compaction. For fast enough 
compaction, the slow cooling regime vanishes.

The averages of the cooling dynamics for all available data from the 
parabolic flight are shown in Fig.~\ref{fig:avg}. For the small compaction 
rate $\Delta\varphi/\Delta t = 0.013/s$, data from parabolas P0, P1 (cf. 
lower panel in Fig.~\ref{fig:cooling}), P2, P3, P4, and P5 are first 
rescaled in time to overlap in the evolution regarding the packing fraction 
$\varphi$ with $\varphi = 0.5$ chosen as $t=0$. Then the data for 
$\Delta(t)$ is averaged over the 6 data sets and shown for the full range of 
pixels as open circles (upper panel of Fig.~\ref{fig:avg}) as well as open 
diamonds (lower panel of Fig.~\ref{fig:avg}). Running averages in time are 
used to obtain the somewhat smoother corresponding full curves. Data for 
compaction rate $\Delta\varphi/\Delta t = 0.017/s$ is treated similarly and 
displayed as filled circles (upper panel) and filled diamonds (lower panel). 
From the averaged dynamics, linear cooling laws can be obtained that are 
consistent with the results from the single runs described above: Compaction 
rate $0.013/s$ is described by $\Delta(t)-\Delta_0 = 10-1.33t$ while 
compaction rate $0.017/s$ follows $\Delta(t)-\Delta_0 = 12.5-t$ in the upper 
panel. The different slopes in those laws follow the variation of the 
overall amplitude of $\Delta(t)$ varies by around 25\%. In the lower panel 
the corresponding laws read $\Delta(t)-\Delta_0 = 2.2(10-1.33t)$ and 
$\Delta(t)-\Delta_0 = 2.2(12.5-t)$, respectively. Hence, the limitation to 
the pixels in the selected region only introduces an additional amplitude.

\begin{figure}[htb]
\includegraphics[width=.95\columnwidth]{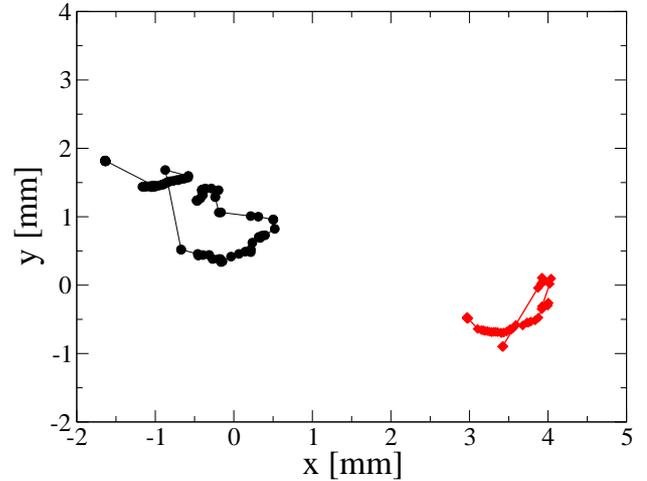}
\caption{\label{fig:trace}Trajectories of two tracer particles 
($d=200\mu$m) inside an assembly of host particles ($d=500\mu$m) during 
compaction in microgravity (parabola 3 on day 3 of DLR-22).
}
\end{figure}

The linear law is valid for around 4s for $\Delta\varphi/\Delta t = 0.017/s$ 
and for 8s for the compaction rate $0.013/s$ which may be accidental. Also 
for the averaged data, the slow linear cooling is followed by a more rapid 
decay of $\Delta(t)$. Again, the final rapid collapse takes place within a 
second and it is observed in Fig.~\ref{fig:avg} that the final decays may be 
scaled on top of each other for different compaction rates. It is possible 
to interpret the data for different compaction rates by a roughly constant 
decay rate $\gamma$ and a shrinking range of validity in time after which 
the final collapse terminates the slow cooling. The fits of the individual 
decay curves for $\Delta(t)$, cf. Fig.~\ref{fig:cooling}, yield such a 
constant $\gamma$ when averaged. It is also possible to imagine that the 
cooling regime vanishes by a decreasing slope $\gamma$ whereby the increased 
energy input at higher compaction rates can overcompensate for the 
dissipation. The latter scenario is consistent with the finding that in the 
fits of the averaged $\Delta(t)$ in Fig.~\ref{fig:avg}, a slight decrease in 
the value of $\gamma$ is obtained. 

While the overall motion can be estimated from the difference images, the 
tracer particles are visible in the transmission images and can be tracked 
individually. Figure~\ref{fig:trace} shows the trajectories of two tracer 
particles during the compaction run shown in Fig.~\ref{fig:P1}. While the 
data are not sufficient to properly define a diffusion coefficient, it is 
seen that despite the relatively high density the tracers travel over 
distances several times their own diameter. Interestingly, the particle 
closer to the center of the cell moves over a larger distance than the 
particle closer to the right piston. However, since the motion in the 
perpendicular direction is not know, no final conclusion can be drawn.

%%%
\section{Stress Birefringence in Three Dimensions}\label{sec:3Dstress}

The first investigation using stress-birefringent particles to model the 
stress transmission in granular packings was done by Dantu \cite[p. 
500ff]{Kuske1974}. In two dimensions, Pyrex glass cylinders were viewed 
between crossed polarizers exhibiting chains of larger stresses when 
penetrated with a piston. In three dimensions, crushed and sieved Pyrex 
glass particles were immersed in an index-matching liquid, allowing a view 
inside the sample also showing stress chains between crossed polarizers. 
Later, the existence of stress chains in three dimensions was also 
demonstrated and analyzed with spherical glass beads 
\cite{liu95,Wood2011}.

\begin{figure}[hbt]\begin{center}
\includegraphics[width=.95\columnwidth]{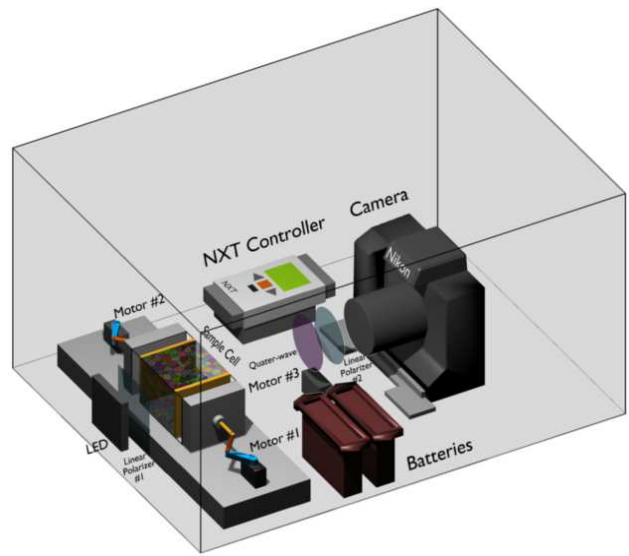}
\caption{\label{fig:setup}Experimental setup for the parabolic flight 
(DLR-13, February 2009) for stress measurements. Stress-birefringent 
particles inside a sample cell with cross section 5cm$\times$5cm and two 
pistons movable by motors 1 and 2 are illuminated by an LED panel from 
behind and recorded between crossed polarizers by a camera. Compression 
and recording is operated by an NXT controller and initiated from outside 
via a bluetooth signal from a cell phone. The entire setup is enclosed in 
an aluminium box, weighs 10kg in total, and is left free-floating for 
distances up to 50cm inside the cabin.
}\end{center}
\end{figure}

While for three dimensions, granular stress-birefringence has so far 
remained largely on the qualitative level, in two dimensions, granular 
stress-birefringence (also called photoelasticity) has been utilized in a 
large variety of instances and analyzed in great detail: In sheared 
systems the transition between loose and load-carrying packings was 
established \cite{howell99}; a qualitative difference in force response to 
external loads was found for ordered and disordered packings 
\cite{geng01}; logarithmic aging of stress was discovered for packings but 
not observed under compression \cite{hartley03}; the relation between 
force chains and friction in stick-slip motion was elaborated 
\cite{Yu2005}; for granular sound, the propagation along force chains was 
demonstrated \cite{Owens2011}; for jamming under isotropic compression, 
non-trivial power-laws were confirmed 
\cite{Majmudar2007,Behringer2008,Zhang2009} and distinguished from jamming 
behavior under shear \cite{Majmudar2005,Utter2008,Zhang2010,Bi2011}.

The investigation of granular stress-birefringence in two dimensions 
continues to be a fruitful route for detailed analysis of the statistical 
properties of granular packings as well as its dynamical properties. 
Motivated by this success, we attempt to extend the methods towards three 
dimensions in the following. Figure~\ref{fig:setup} shows the setup of the 
granular compaction experiment for the parabolic flight campaign DLR-13 in 
2009. The sample cell containing the granular particles is illuminated 
from behind by an LED panel which has a polarizer laminated on top of it. 
A rotatable quarter-wave plate in addition to a polarizer in front of a 
camera (Nikon D3) completes the polariscope. Pistons at two walls of the 
sample cell can be moved by servo motors to change the volume and allow 
for the compaction of the granular particles from a loose assembly into a 
force-carrying packing. The switching of illumination, the motion of the 
motors, the rotation of the quarter-wave plate, and the multiple release 
of the camera is fully automated by an NXT controller. The experiment box 
is left floating freely for up to 50cm inside the plane. Measurements are 
performed during the 0g-phase of the parabolic flight and in order to 
minimize the disturbance, the start of the measurement is triggered from 
the outside by a cell phone (Nokia 6131) via bluetooth.

\begin{figure}[hbt]\begin{center}
\includegraphics[width=.75\columnwidth]{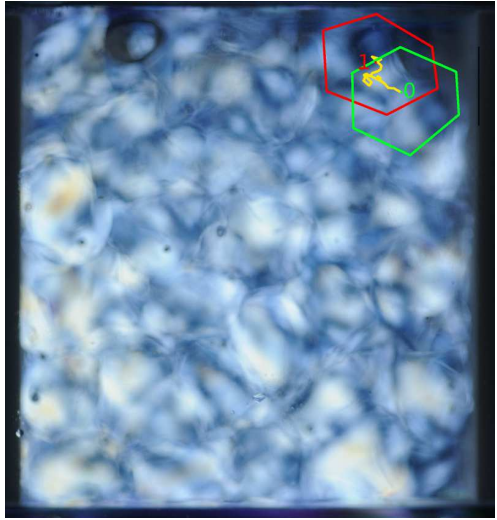}
\caption{\label{fig:rattler}Rattler motion observed against the 
force-network backbone visualized by stress-birefringence. The outlines 
indicate the rattler at its initial (0, green) and end (1, red) position. 
The path (yellow) shows the distance traveled.
}\end{center}
\end{figure}

Since glass has a very small stress-optical coefficient, high external 
pressures (of the order of several 100kPa) are needed to obtain an 
appreciable signal from a packing of glass particles. In contrast, the 
present experiments are performed with gelatine particles in a 
water-glycerol mixture for index-matching. While gelatine has been used 
for photoelastic investigations (see \cite{Kuske1974} for a review) the 
production of stress-free particles needed to be refined and shall be 
detailed elsewhere. The use of gelatine reduces the demand on the 
mechanical structures of the microgravity device and hence allows for 
easier implementation. Already below 100Pa, an assembly of gelatine 
particles exhibits a reliable signal. In addition, one can approach much 
closer the transition point where the granular particles lose or establish 
contacts. In order to prevent ordering for particles of the same size, 
irregular particles are cast with a mean diameter of around 9mm with a 
shape indicated by the outlines in Fig.~\ref{fig:rattler}.

The state of a partially index-matched sample of irregular particles is 
shown in Fig.~\ref{fig:rattler} after compaction from left and right with 
the pistons of cross section 5cm$\times$5cm. The corresponding motion of 
the ratter particle after compaction in microgravity is demonstrated in 
the movie 0g-rt.mpg in the supplementary material. The color fringes 
reveal the existence and inhomogeneity of the stresses inside the force 
network in the sample. The dark sphere at the upper left end of the 
picture is an air bubble. The partial index match allows for the 
simultaneous observation of particle motion and it is found that among 
around 300 particles only a single particle in the front upper right 
corner is still moving. The trajectory of the particle's center is similar 
to the motion of the tracers in the X-ray experiment, cf. 
Fig.~\ref{fig:P1}. The distance traveled by the rattler is about half its 
diameter. The differences in the shapes of the outline in the beginning 
and the end of the trajectory are due to the rotation of the rattler in 
its pocket formed by the arrested particles.

%%%
\section{Conclusion}\label{sec:conc}

It has been shown that X-ray radiography and stress-birefrin\-gence allow 
the observation of the compaction of a granular packing in microgravity. 
Remarkably, the conditions on parabolic flights are especially suitable to 
observe rattlers that are agitated by the rest-accelerations without 
destroying the packings. Both methods can identify reliably the motion of a 
small fraction of rattler particles among the network of particles that form 
the backbone of the packing. While not enough data is currently available 
for an elaborate analysis of rattler dynamics from 3D stress-birefringence 
or the tracer dynamics in X-ray radiography, the results show nevertheless 
that microgravity experiments give access to new phenomena not observable on 
ground.

For the X-ray radiography data it is possible to quantify the bulk dynamics 
in the samples, resulting in much more reliable statistics. Using the time 
gradient by analyzing the difference images from the detector, a reliable 
quantity $\Delta(t)$ can be obtained to characterize the motion of the 
particles. $\Delta(t)$ allows the distinction between agitated and arrested 
states. In addition, it is possible to identify a novel regime of cooling 
quantitatively for low rates of compaction. This is only possible in 
microgravity as under the dominating influence of gravity granular gases 
collapse quite rapidly \cite{Son2008}. The newly identified cooling extends 
over several seconds and is described reasonably well by a linear decay of 
$\Delta(t)$.

\begin{acknowledgements}
We acknowledge financial support by DFG FG1394 and BMWi 50WM0741 as well as 
technical assistance by D. Br\"auer, F. Kargl, S. Klein, and T. Kornwebel. 
PY and MS want to thank especially warmly their advisor and tutor Bob 
Behringer for fruitful guidance over many years. 
\end{acknowledgements}

\bibliographystyle{apsrev}
\bibliography{lit,addlocal}
\end{document}